\documentclass[12pt,preprint]{aastex}

\usepackage{graphicx}
\usepackage{times,psfig}

\newcommand\lsim{\mathrel{\spose{\lower 3pt\hbox{$\mathchar"218$}}
     \raise 2.0pt\hbox{$\mathchar"13C$}}}
\newcommand\gsim{\mathrel{\spose{\lower 3pt\hbox{$\mathchar"218$}}
     \raise 2.0pt\hbox{$\mathchar"13E$}}}
\def\ltsima{$\; \buildrel < \over \sim \;$}
\def\lsim{\lower.5ex\hbox{\ltsima}}
\def\gtsima{$\; \buildrel > \over \sim \;$}
\def\gsim{\lower.5ex\hbox{\gtsima}}

\def\apj{ApJ}

\def\apjs{ApJSS}

\shorttitle{Spectroscopy of SDSS J004054.65-0915268}
\shortauthors{Landoni \& Zanutta et al.}

\begin{document}


\title{Optical spectroscopy of SDSS J004054.65-0915268: three possible scenarios for the classification.  A $z \sim 5$ BL Lacertae, blue FSRQ or a weak emission line quasar.}


\author{M. Landoni \& A. Zanutta, A. Bianco, F. Tavecchio, G. Bonnoli and G. Ghisellini}
 \affil{INAF-Osservatorio Astronomico di Brera, Via Emilio Bianchi 46, I-23807 Merate, Italy}

\begin{abstract}
The haunt of high redshift BL Lacerate objects is day by day more compelling, to firmly understand their intrinsic nature and evolution. SDSS J004054.65-0915268 is, at the moment, one of the most distant BL Lac candidate at $z\sim5$ \citep{plot10}. We present a new optical-near IR spectrum obtained with ALFOSC-NOT with a new, custom designed dispersive grating aimed to detect broad emission lines that could disprove this classification. In the obtained spectra we do not detect any emission features and we provide an upper limit to the luminosity of the C IV broad emission line. Therefore, the nature of the object is then discussed, building the overall spectral energy distribution and fitting it with three different models. Our fits, based on the SED modeling with different possible scenarios, cannot rule out the possibility that this source is indeed a BL Lac object although, the absence of optical variability and lack of strong radio flux, they seems to suggest that the observed optical emission originate from a thermalized accretion disk.\end{abstract}





\section{Introduction}

BL Lac objects are active galactic nuclei dominated by a strongly beamed nuclear non thermal emission arising in a relativistic jet pointed toward the Earth (e.g., \cite{urry95}). Relativistic effects amplify the radiation produced in the jet by relativistic electrons through synchrotron and inverse Compton mechanisms. Moreover, contrary to flat spectrum radio-quasar, also the intrinsic nuclear thermal components are quite weak (e.g. \cite{ghisellini11}). As a consequence, the non-thermal continuum out-shines spectral features (emissions of absorptions), making the determination of their redshift quite tricky \citep{landfors, landxsh} . Nevertheless, the estimate of their distance is obviously important to shade the light on the nature of their radiation mechanism and about their evolution.
In fact, radio selected BL Lacs display moderate positive evolution, while X-ray selected objects negatively evolve when derived from X-rays survey, or not evolve at all \citep{giommi09, giommi12, pad07, rector2000}. A continuum trend from slightly positive-evolution for low energy peaked BL Lacs, to strong negative-evolution in the case of high peaked ones was proposed \citep{rector2000}. However, the statistics concerning the evolution of the BL Lacs suffer of redshift incompleteness, making the increase of the objects with reliable redshifts (from homogeneous and unbiased selections) a core issue (see also \cite{shaw} for a discussion). A more intriguing fact is that these objects are detected almost in the low-z Universe (z $\leq$ 1) and very few sources are known at z $\geq$ 2 (see e.g. \cite{shaw}). 
This has been interpreted in terms of the evolution from highly accreting sources (FSRQ) to BL Lacs, in which accretion is rather low \citep{cavaliere,bott}.

For all of these reasons, the search of high-redshift BL Lacs is becoming even more compelling.
With this aim, we searched in the most extensive BL Lac catalogs to find high-$z$ candidate sources, focusing on the compilation of the homogeneuous optically selected BL Lac objects from Sloan Digital Sky Survey by \cite{plot10}. We found that, in the sample of radio loud BL Lac candidates, the source SDSS J004054.65-0915268 (hereafter 0040--0915)  is characterised by a featureless optical continuum and its redshift, based on an Lyman limit system (LLS), is surprisingly high (z$ = 4.976$), making it one of the most distant BL Lac candidates. It is worth to mention that \cite{fan06} serendipitously found another, event more distant, candidate BL Lac object (but see \cite{lei} classifying it as QSO). However, the required spectral range to reveal broad emission lines redward of the Ly-$\alpha$ line, such as the C IV, is unfortunately outside the sensitivity range of our instrument in this object. For this reason we thus select 0040--0915 for our spectroscopic investigation whose nature is matter of debate in literature.  In fact, while \cite{collinge} and \cite{plot10} consider the source as BL Lac candidate (see Table 5 in both papers), \cite{lane11} on the basis of Spitzer mid-IR data suggest that the source is not dominated by non thermal continuum, challenging the BL Lac hypotesis. This source has been considered as Weak Emission Line Quasar (WELQ), a new recent class of AGN with weak or absent broad emission lines, by \cite{dia09}. Despite having only an upper limit in the radio band it was classified as a Radio Loud BL Lac Candidate object by \cite{plot10}.  The different classifications of this source rely mainly on the SDSS spectrum which, unfortunately, can determine only the presence or absence of the broad Lyman-$\alpha$ line. This spectral feature could be weak for the combination of both severe intervening absorptions or poor SNR in that band. 

In the SDSS spectrum of 0040--0915, the Ly-$\alpha$ emission line appears to be absent while other weak features could be interpreted as S IV or N V emission. We note, however, that a proper modelling of the continuum and emission line should be used to robustly infer the detection of the feature. However, the poor signal to noise ratio of the SDSS spectrum does not allow to further speculate on those emission lines. The C IV broad emission line, whose detection allows to disentangle between scenarios, could be only partially detected nearly the end of the spectral range (in the near-IR band). For all these reasons, a good quality near IR spectrum largely covering the region of the expected C IV feature (up to 1$\mu$m) is mandatory to possibly firmly detect the line and clearing the fog on the scenario between normal broad line QSO, BL Lac or exotic source.

In order to fill this gap, we exploited the upgrade of dispersive element at Andalucia Faint Object Spectrograph and Camera (ALFOSC) spectrograph at Nordic Optical Telescope (NOT, La Palma, ES) designing a new red grating to match these particular requirements aiming to perform the needed observation. Although in the literature there are a number of papers that discuss the existence of quasars with weak emission lines (see, e.g. \cite{dia09, plot10, lane11,meu14}), no exhaustive broad band Spectral Energy Distribution (SED) are interpreted adopting various models, ranging from normal QSO to BL Lacs. In particular, in this work we present SEDs assembled with data from far-IR to X-Rays data and we apply physical models to possibly shade light on the nature of the source.  In this paper, we thus show both the obtained spectroscopic results along with three possible frameworks for the interpretation of this high redshift source by the adoption of popular scenarios (see, e.g. \cite{madau,maraschi}). 

The paper is organised as follow. We report in Section 2 the grating manufacturing and the data reduction while main results are reported in Section 3. Discussion and conclusions are given in the last section. We use cgs units unless stated otherwise and the following cosmological parameters $H_{0} = 70$ km s$^{-1}$    Mpc$^{-1}$, $\Omega_{m} = 0.27$, $\Omega_{\Lambda} = 0.73$.

\section{Grating manufacturing, observations and data reduction}

The dispersive element used for the spectroscopic acquisition highlighted in this work was specifically designed and produced in the INAF (OABr, Merate) facilities.
The possibility to design and manufacture a dispersive element specific for a target observation, is surely effective for addressing the future astronomical scientific cases. 
The logic behind this new custom device was to obtain a low dispersive element with an extended wavelength range, which is capable to detect the C IV broad band line of the object 0040--0915.
This device, called Grism \#20, was integrated at NOT-ALFOSC and its specifications are reported in Table \ref{tab:GR20}.

\begin{table}[h]
\caption{Grism \#20 main specifications' summary. 
Col. (1): working central wavelength of the grating; Col. (2): device's 1--st order diffracted efficiency; 
Col. (3): pixel dispersion of the grism; Col. (4): wavelength range; Col. (5): calculated resolution; Col. (6): 
pitch of the grating.}
\label{tab:GR20}
\begin{center}
\begin{tabular}{ |c|c|c|c|c|c|c| }
\hline
\rule[-1ex]{0pt}{3.5ex} GRISM & $\lambda_{central}$ & efficiency & dispersion & $\Delta \lambda$ 
&resolution & l/mm  \\
\rule[-1ex]{0pt}{3.5ex} name & [\AA] (1) & at 7350 \AA (2) & [\AA / px] (3) & [\AA] (4) & (5) & (6) \\
\hline
\hline
\rule[-1ex]{0pt}{3.5ex} Grism \#20 & 7350 & 88\% & 2.079 & 5760 - 10200 & 895 & 484 \\
\hline
\end{tabular}
\end{center}
\end{table}

This dispersive element is a grism (GRating $+$ prISM), based on a Volume Phase Holographic Grating (VPHG) and manufactured with state of the art photopolymeric material (Bayfol$^{\circledR}$ HX), developed by Bayer MaterialScience AG \citep{bayfol2013}.
Such photopolymers show an easy processability (they are self-developing) together with a large modulation of the refractive index ($\Delta$n) and, even more interesting, the $\Delta$n can be finely tuned by acting on the writing conditions. These materials are becoming a reliable alternative to dichromated gelatin (DCG, the reference material for VPHGs), with high dynamic range and sensitivity \citep{zanuttaPASP, zanuttaSPIE}.
\\
We delivered the grating at ALFOSC-NOT on 27 September 2014 and we observed 0040--0915 during the night of 30 September and 1 October 2014. We secured two individual exposures of 2200s each using the 1.8$^{\prime\prime}$ slit. The seeing during the night was $\sim$ 1.50$^{\prime\prime}$ and the sky was almost clear. We also observed two standard stars with the same configuration of the instrument to flux calibrate the two spectra. Standard IRAF\footnote{
IRAF (Image Reduction and Analysis Facility) is distributed by the National Optical Astronomy Observatories,
which are operated by the Association of Universities for
Research in Astronomy, Inc., under cooperative agreement
with the National Science Foundation.} procedures were adopted (bias, flat field correction, wavelength and flux calibration) to fully reduce each observation. Finally, we averaged the two spectra according to their signal-to-noise ratio (S/N $\sim$ 15) in order to obtain the mono dimensional spectrum.  We also cross checked the absolute flux calibration through an aperture photometry of the i-band acquisition image of the field, secured just before obtaining the spectra.

\section{Results}

			\begin{figure}[h!tb]
				\centerline {\includegraphics[scale=0.75]{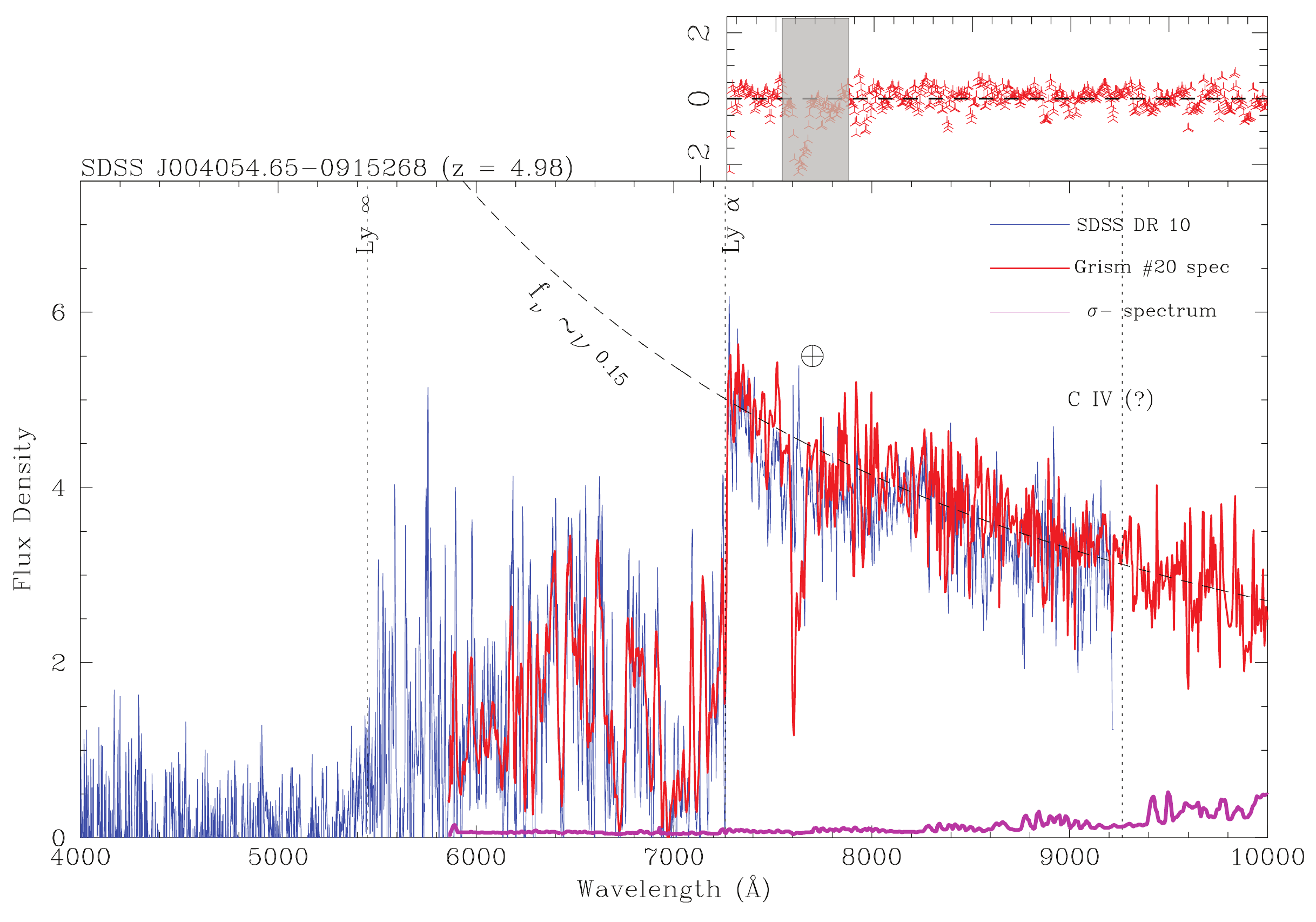}} 
\caption{
SDSS J004054.65-0915268 optical spectrum in the band 4000 $\textrm{\AA}$ - 1 $\mu$m. Units are in 10$^{-17}$ erg cm$^{-2}$ s$^{-1}$ $\textrm{\AA}$$^{-1}$. The blue solid line is the spectrum available in the Sloan Digital Sky Survey DR10 while the red solid line represents the newly acquired spectrum at ALFOSC-NOT. Telluric absorption lines are marked with $\oplus$. The redshift of the source, estimated through Ly-$\alpha$ break and Lyman-limit features, is $z = 4.981$. The presence of broad emission lines ascribed to Ly-$\alpha$ and C IV is ruled out up to EW $\sim$ 5 $\textrm{\AA}$. The dashed black line is the continuum best fit ($F(\nu) \propto \nu^{0.15}$). The purple solid line is the 1-$\sigma$ spectrum. Upper right panel: residual between the fit and the observed NOT spectrum. 
} 
\label{fig1}
\end{figure}

We report in Figure \ref{fig1} the optical spectrum of 0040--0915 obtained by the SDSS DR10 (blue solid line) and the newly one secured with ALFOSC-NOT equipped with Grism \# 20 (red solid line, SNR $\sim$ 15). The redshift of the source, estimated through the Ly-$\alpha$ brake and the Lyman-limit, is $z \sim 4.981$ and no intrinsic broad emission lines (neither Ly-$\alpha$ nor C IV) are detected.  We mark in Figure \ref{fig1} the expected position of the C IV emission line at redshift mentioned above ($\lambda \sim$ 9300)  with a vertical dotted line. We fitted the continuum by the adoption of a power law of the form f$_\nu$ $\propto$ $\nu^{0.15}$, consistent with the results obtained by \cite{fan06} for a similar a high-z lineless AGN (see Figure 3 of \cite{fan06}). Residuals are reported in the right-top panel of Figure \ref{fig1}. The SDSS spectrum, due to its poor S-N ratio at the edge of spectral sensitivity and its short range in the near-IR (see Figure \ref{fig1}) cannot rule out the presence of a broad emission line ascribed as C IV . 
We note, however, that the SDSS best fit reported in DR10 could have been tumbled a possible weak emission of Ly-$\alpha$ although in that region the fit is rather poor and systematically underestimate the continuum. This hint is not, however, detected in our newly obtained spectra (see Figure \ref{fig1}). In fact, the fit adopted by the SDSS could mimic the presence of spectral features that are not obiouvsly present in the data.  For instance, the detection of S IV emission line has been ruled out in the newly DR10 release.
\subsection{CIV equivalent width upper limit}
\begin{figure}[h!tb]
 \includegraphics[scale=1.35]{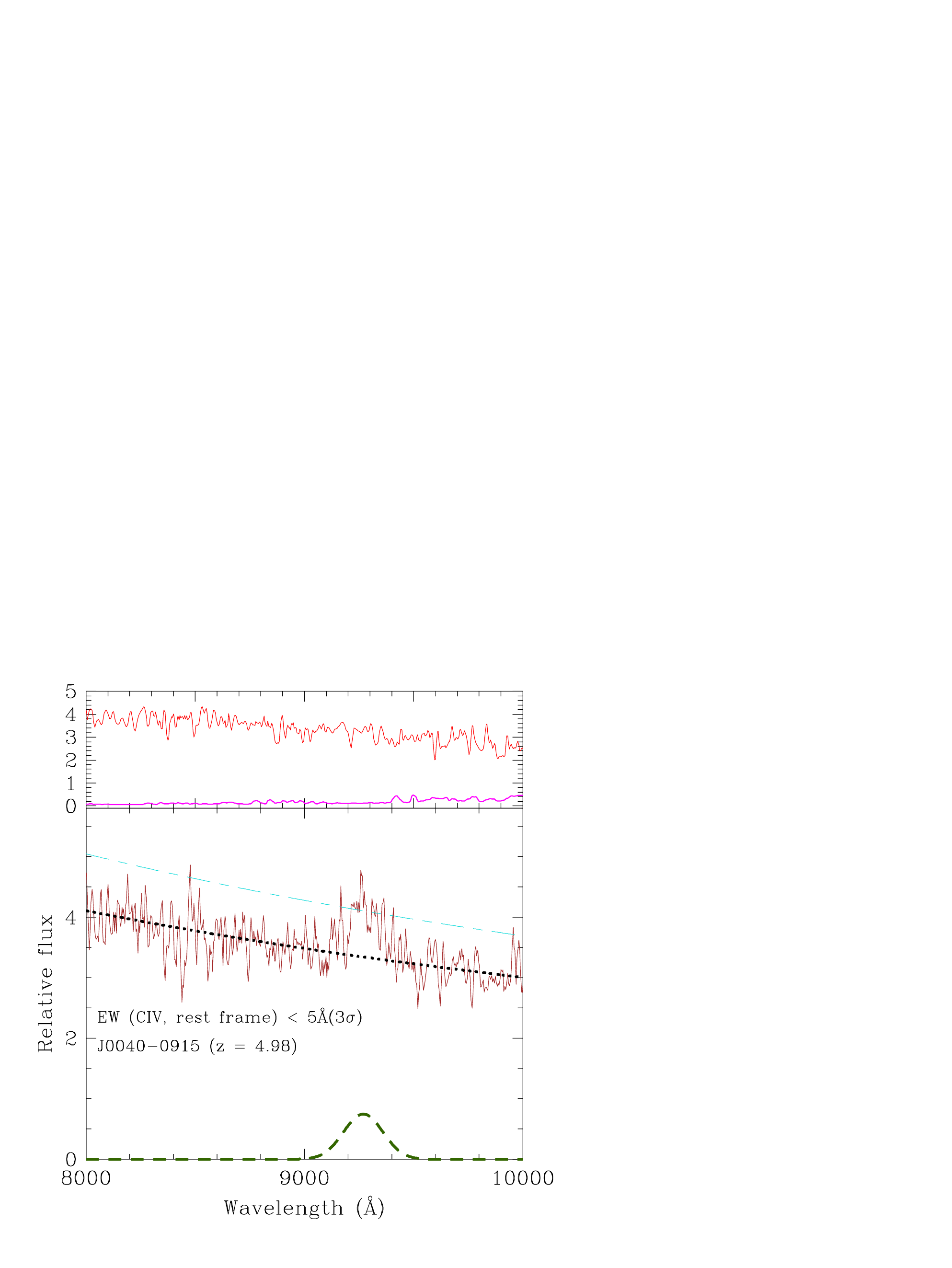}
\caption{
C IV emission line equivalent width limit estimation. In the bottom panel is reported the simulated spectrum by the adoption of the power-law (dotted black line, estimated on the NOT spectrum) and a C IV emission line model with EW of 5 Angstrom  rest frame (dark green dashed line) and a FWHM of 4000 km s$^{-1}$ rest frame. The resulting spectrum, which is the sum of the component plus a noise to simulate the same SNR of the NOT spectrum, is indicated by the solid brown line. The $3-\sigma$ threshold is indicated by light blue dashed line. Upper panel: Comparison of the observed NOT spectrum (solid red) and the 1$-\sigma$ spectrum (solid purple line). Details on the estimation are given in the paper. A coloured version of the figure is available on the online edition of the Journal.
} 
\label{fig2-limit}
\end{figure}
The ALFOSC NOT spectrum, which extends up to 1 $\mu$m, has a sufficient spectral coverage and decent S-N ratio to safely exclude the presence of C IV broad emission line (see Figure 1). Following the description by \cite{fan06} we modeled the spectrum in region 8000-10000 (where the S-N ratio and resolution can be considered roughly constant for our estimation) using the power-law described before plus an emission line model (FWHM $\sim$ 3000-4000 km s$^{-1}$) for the CIV and applying to the resulting model a S-N ratio similar to that observed in the NOT spectrum. We performed simulations by increasing the EW of the line until the signal in the simulated spectrum exceed the noise by 3-$\sigma$. We repeated this recipe 100 times, as in a usual Monte Carlo simulation. 
\\We report in Figure \ref{fig2-limit} results obtained from this procedure. We show  in the lower panel the two components of the model (dotted line for the power-law and dashed line for the line) and the resulting spectrum (brown solid line) after applying gaussian noise to simulate our observed SNR ratio. Since the C IV emission line has a rest frame EW of 5$\textrm{\AA}$ and its signal exceed the noise with a threshold level of 3-$\sigma$ we can conclude that the upper limit to the EW of CIV is 5$\textrm{\AA}$ at 3-$\sigma$ confidence level. For comparison, we report in the upper panel the observed NOT spectrum (red) and the 1-$\sigma$ error spectrum (purple solid line). As a sanity check, we also calculated the minimum equivalent width in the same region following the procedure described in \cite{sba06} and adopted in various papers to assess limits on the detectability of spectral features (see e.g. \cite{sch93,sba06,farina}). This method is completely model independent and relies only on the width of the spectral element (which is a function of the instrument) and the S-N ratio (see \cite{sba06} for details). In particular, we evaluated the EW on bins of the size of the resolution element in various regions of the spectrum (excluding the telluric structures) and we assume as EW$_{min}$ the 3-$\sigma$ deviation from the mean ($\mu \sim 0 \textrm{\AA}$) of the distribution of the EWs obtained in each bin. We found that the EW$_{min}$ (rest frame) is $\sim$ 5 $\textrm{\AA}$ consistent with the limit obtained with the model described above.
We can thus conclude that, if the C IV feature is present, shall have an Equivalent Width (EW)  $\leq$ 5 $\textrm{\AA}$ rest frame. We report in the following section the possible scenarios that could explain the observed optical-nearIR spectrum after discussion on caveats on the estimation of the redshift of the source. 
\subsection{Caveats on the redshift of 0040-0915}
\begin{figure}[h!tb]
 \includegraphics[scale=1.35]{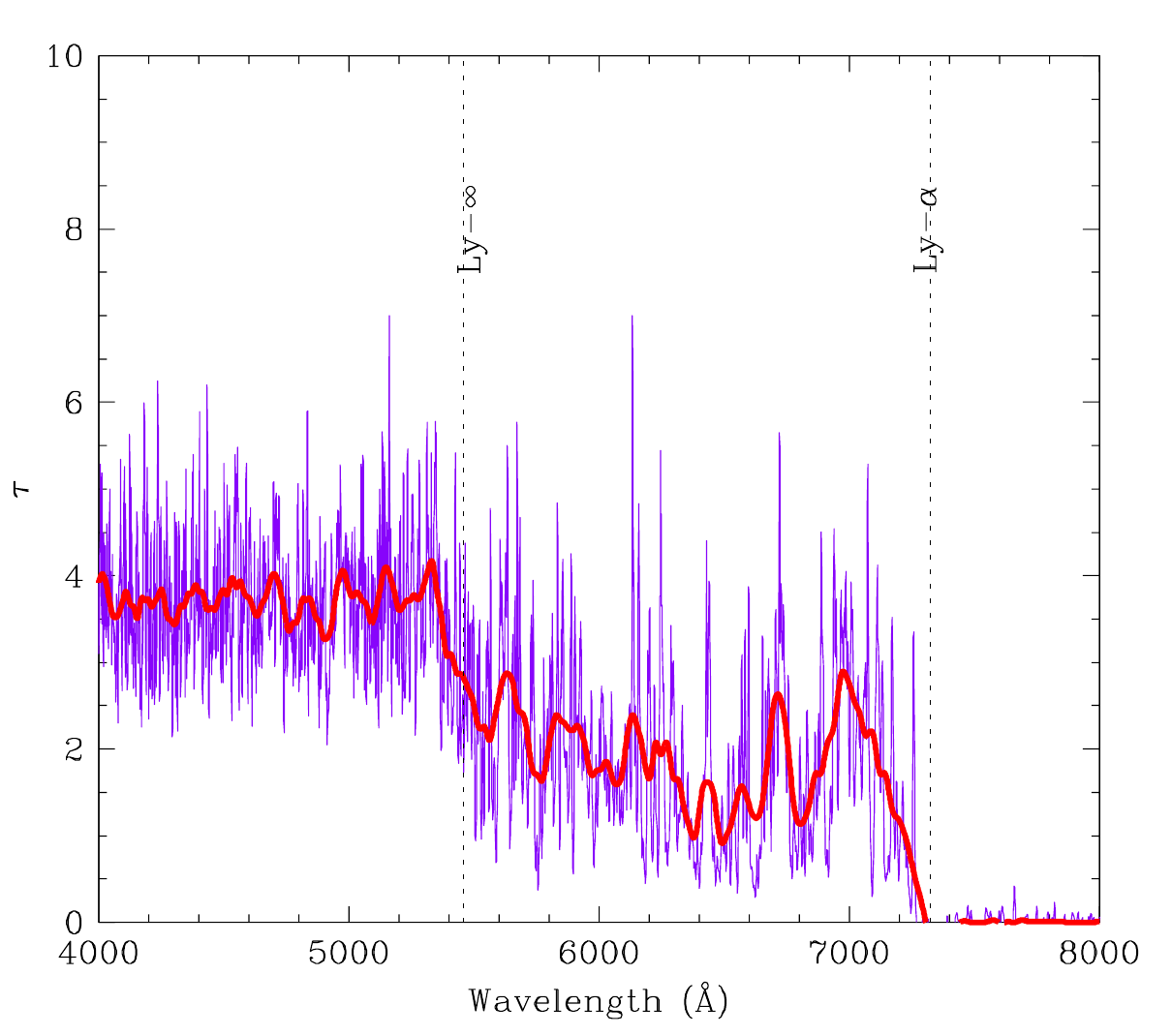}
\caption{
Optical depth derived from the spectrum of 0040--0915. The blue solid line is the ratio between the observed flux on the SDSS spectrum and the intrinsic one extrapolated from the power-law fit (see Section 3.2). The thick solid line is 20 $\textrm{\AA}$ boxcar filter on the blue spectrum. The redshift of the nearest H I cloud is assessed through the wavelength where the optical depth drops to $\tau \sim 0$. This value catches the cold H I cloud nearest to 0040--915. The expected position of the Lyman-$\infty$ is consistent with our interpretation (see text of Section 3.2) and it is marked with a vertical dotted line. 
} 
\label{fig3-tau}
\end{figure}
The redshift estimation from the onset of Lyman-$\alpha$ forest can be biased to higher values (up to $\Delta z \sim$0.1) if the EW of the broad Lyman-$\alpha$ line is very small (see e.g. Figure 3 of \cite{dia09}). Although in the NOT and SDSS spectra the continuum appears to be dominated by the power-law component and this effect could be mitigated, other biases should be considered when redshift is derived only through absorption features (see e.g. \cite{prochaska2008}  and Figure 1 and 2 in \cite{steidel2010}). In particular, the onset of the Lyman-$\alpha$ forest can be affected by absorption both blueward and redward of the Lyman alpha emission line. When considering systems like proximate Damped Lyman Alpha System (Prochaska et al. 2008) absorptions near the systemic velocity that completely absorbs the center of the Lyman-$\alpha$ line could move the onset of the Lyman forest towards redder wavelengths. Overall, these effects can bias the observed redshift up to $\Delta z \sim$ 0.01.
\\

We tried to check the redshift of 0040--0915 through an alternative way based on the optical depth. Following \cite{fan06} we defined the optical depth as $\tau = -\log \frac{f_{obs} }{f_{int} }$  where f$_{obs}$ is the observed flux of the source and f$_{int}$ is the extrapolated intrinsic flux obtained from the power law fit. We report the results (using the SDSS spectrum) in Figure 3. The wavelength location of $\tau = 0$ marks the 1216$\textrm{\AA}$ features which corresponds to $z = 4.989$. This value is consistent with the one considered in this paper and confirmed in the SDSS DR10 spectrum analysis ($z = 5.002$). Consistently, we also mark the position of the Lyman edge at 912$\textrm{\AA}$ at that redshift. As expected, at that location the optical depth monotonically increases up to the the saturation plateux at $\tau \sim 4$ (see e.g. \cite{peterson}). 
\clearpage

\section{Discussion}

The rest frame near--IR UV spectral energy distribution (SED) of 0040--0915 is shown in Figure \ref{zoom}, reporting archival data\footnote{from ASDC: \url{http://www.asdc.asi.it}, SDSS and Spitzer (see \cite{lane11})} and our spectrum, and in Figure \ref{combo}, reporting the overall SED compared with three different models.

Weak or absent broad emission lines is the main defining property of BL Lac objects, namely jetted sources whose jet is pointing at us and whose jet emission is strongly boosted by the bulk relativistic motion of the emitting plasma.
On the other hand, {\it all} known BL Lac objects are radio--emitters, as expected in the case of a jetted source. The other property of BL Lac objects pertinent to our case is the absence of visible accretion disc emission. Our source, instead, has a IR--optical SED showing a peak (in $\nu L_\nu$) at a rest frame frequency in the UV band, and a rather hard spectrum below. The classification of 0040--0915 is therefore a puzzle. In the following we interpret 0040--0915 alternatively as a radio--quiet quasar, a flat spectrum radio quasar (FSRQ) with a particularly enhanced and blue jet emission, and a BL Lac object.

The adopted model for the disc and jet emission is described in Ghisellini \& Tavecchio (2009). We summarise here its main features and parameters. The disc is a standard Shakura--Sunyaev (1973) disk, with an X--ray corona phenomenologically described as emitting a cut--off power law spectrum: $F_{\rm corona}(\nu)\propto \nu^{-\alpha_X} \exp(-\nu/\nu_{\rm c})$. The non--thermal source is assumed to be a sphere located in the jet at some distance $R_{\rm diss}$ from the black hole. It is embedded in a homogeneous and tangled magnetic field $B$ and moving with a nulk Lorentz factor $\Gamma$ at a viewing angle $\theta_{\rm v}$.
Throughout the source relativistic electron are injected, isotropically in the rest frame, with a power $P^\prime_{\rm e}$ as measured in the comoving frame. They are injected with a smoothly joining broken power law, with a break at a random Lorenz factor $\gamma_{\rm b}$, up to the maximum random Lorentz factor $\gamma_{\rm max}$. They produce synchrotron and inverse Compton radiation. The seed photons for scattering are the ones produced internally to the jet, by the synchrotron process, and the ones produced externally to the jet, by the accretion disc, its X--ray corona, the BLR and the torus.
The jet power is in the form of produced radiation ($P_{\rm r}$), Poynting flux ($P_{\rm B}$), kinetic energy of the emitting electrons ($P_{\rm e}$) and protons ($P_{\rm p}$), assumed to be cold and one per each emitting electron.

We remark that our aim in reproducing the SED of the source is not to find out the exact values of the parameters but, using typical values for the parameters (see e.g Ghisellini et al 2010), check the consistency of the different classifications.

\subsection{The radio-quiet quasar scenario}
Figure \ref{zoom} shows the spectrum produced by a disc around a black hole of mass $M_{\rm BH} =1.9\times 10^9 M_\odot$ accreting at the 45\% of the Eddington rate. This interpretation is consistent with the existing data, considering that the emission of the torus (contributing to the IR luminosity) is rather uncertain. We can then interpret 0040--0915 as radio--quiet quasar, with a large black hole mass and accreting vigorously.
But unlike other quasars with similar accretion rates and black hole masses, in 0040--0915 there is no (or very weak) broad emission line region. The SED in Figure \ref{zoom} shows that there is no deficit of ionising photons, so the absence of the BLR must be caused by another reason.

\subsection{The radio-loud scenarios}
An alternative possibility is to detect the source in deep radio band observation with flux just below the current upper limit. In fact, in this case the source would be characterised by a radio loudness $R_L \geq 15$, that, according to the classical definition by \cite{kel}, allows to classify the source as radio loud. This however opens up the possibility of two subcases. The first one (i) the jet non-thermal continuum contributes to the optical-UV emission partly swamping the emission lines or (ii) the jet is intrinsically weak jet or misaligned and therefore does not contribute to the optical-UV continuum. In the latter case this does not solve the problem of weak emission lines, as for the radio quiet case. Therefore, we concentrate on the first case presenting two canonical examples. 
\subsubsection{The Blue FSRQ scenario}
We next investigate the case of a FSRQ with a jet pointing at us and emitting a non--thermal spectrum able to hide the lines, of normal luminosities. This requires that the jet continuum also hides the radiation produced by the accretion disc. To have an estimate or upper limit to the disc emission in this case, we have used our IR--optical spectrum to derive an upper limit of the Ly$\alpha$ and  CIV broad lines,
resulting in $L_{{\rm Ly}\alpha}\lsim 1.1 \times 10^{45}$  erg s$^{-1}$ and $L_{\rm CIV} < 7\times 10^{44}$ erg s$^{-1}$.
Assuming the template in Francis et al. (1991; see also Vanden Berck et al. 2001) we can derive an upper limit for the disc luminosity of $L_{\rm d}< 6\times 10^{46}$ erg s$^{-1}$. Note that our IR--optical spectrum and archival data correspond to a luminosity exceeding $10^{47}$ erg s$^{-1}$.
The main difficulty of this scenario is the very poor fit to the optical (i.e. rest frame UV) data, severely underestimated.
The blue appearance of the jet continuum is achieved by assuming that the dissipation region $R_{\rm diss}$ of the jet is beyond the broad line region  $R_{\rm BLR}$. This implies a deficit of seed photons (produced externally to the jet) that would otherwise rapidly cool the emitting electrons, shifting their relevant energies towards smaller values, and thus producing a redder spectrum (i.e. a synchrotron flux peaking in the sub--mm, and a Inverse Compton flux peaking at $\sim$10 MeV, as is the case for ``normal" FSRQ (see, e.g. Ghisellini \& Tavecchio 2015).
 
\subsubsection{The BL Lacertae scenario}
\begin{figure}[h!tb]
				\centerline {\includegraphics[scale=0.7]{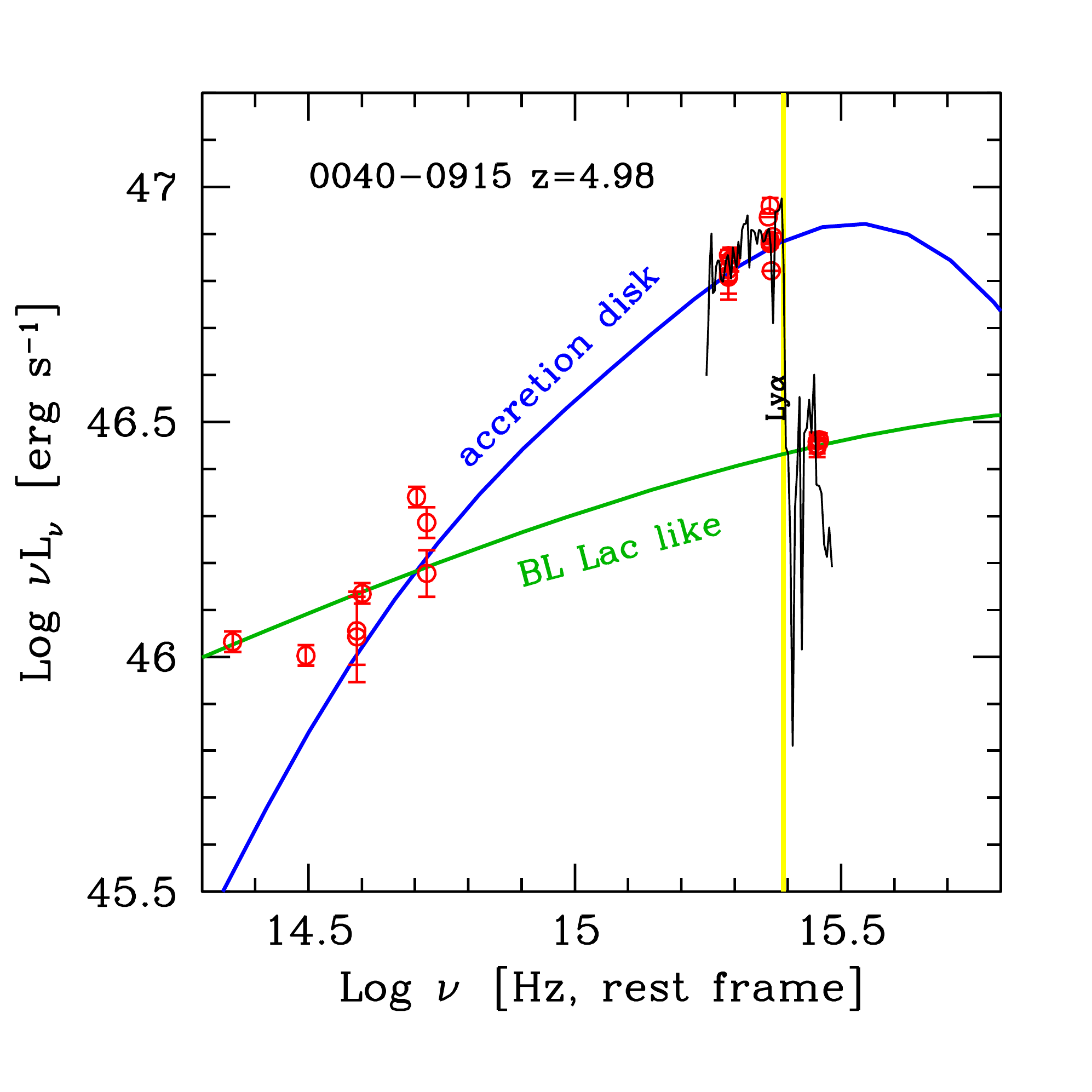}} 
\caption{
The IR--optical SED of 0040--0915 in the rest frame. The archival photometry data (from SDSS, WISE and Spitzer/IRAC) are red circles while our optical spectrum is the solid black line. The vertical yellow line indicates the hydrogen Ly$\alpha$ frequency. Above this line we expect absorption due to intervening matter.
The SED is compared with two different models: the solid blue line is the spectrum from a standard Shakura \& Sunjaev (1973) disc, while the solid green line is the non thermal continuum produced by a relativistic jet that fits the overall SED (as illustrated in Figure \ref{combo} by the mid and bottom panels).
} 
\label{zoom}
\end{figure}

\begin{figure}[h!tb]
				\centerline {\includegraphics[scale=0.65]{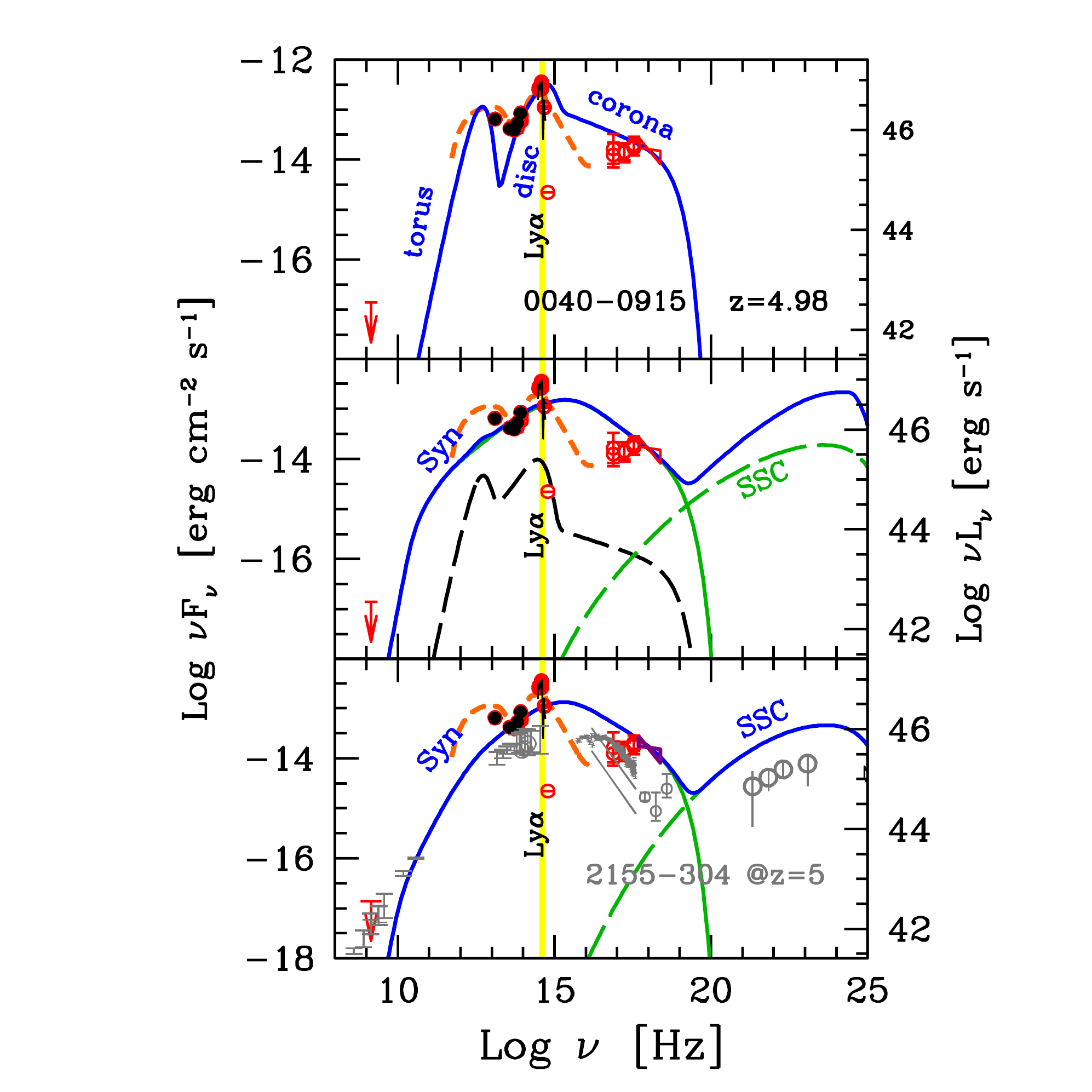}}
				\caption{Spectral energy distribution of 0040--0915 interpreted with three different models.
In all panels the vertical yellow stripe indicates the location of the hydrogen Ly$\alpha$ line. Top panel: the source is assumed to be a radio--quiet quasars, whose accretion disc, torus and X--ray corona are responsible for all data points. 
In this case the anomaly of the source is the presence of a strong ionising continuum produced by the disk without the corresponding presence of a ``normal" broad line region. Mid panel: the source is assumed to be a ``blue quasar", namely a radio loud FSRQ whose non thermal jet emission overwhelms the thermal disc radiation.
This implies a rather ``blue" jet spectrum (Ghisellini et al. 2012).
In this case the broad lines, although present, are hidden by the jet radiation.
The bottom panel illustrates the case of a BL Lac object, in which the broad lines are not produced because of the very weak radiation produced buy the disc (that likely is in the inefficient ADAF regime).
We compare the spectrum of 0040--0915 with a well known BL Lac object, PKS 2155--304 as it would be seen if it were at $z=4.98$.
Note that the radio emission of PKS 2155--304 (if it were at $z=4.98$) is consistent with the upper limit, and that the overall energetics and type of spectrum are similar. Both in this and in the ``blue quasar" interpretations, the problem is the very poor agreement with the IR--optical data. For comparison the SED of QSO is reported (orange dashed line, \cite{rich}). Black dots are from Spitzer IR  (Lane et. al. 2011).}
				\label{combo}
			\end{figure}

A second possibility to account for a moderate radio loud source without strong emission lines, is that the source is a BL Lacertae object as tentatively classified by \cite{plot10}. If 0040--0915 is a BL Lac object, then the disc emission would be negligible, and the absence of ionising photons would be the cause of the absence of the broad emission lines.
The entire emission would be produced by the jet.
As can be seen in Figure \ref{zoom} and Figure \ref{combo}, the main difficulty with this interpretation is the poor fit to the optical data and the infrared Spitzer data \citep{lane11}, as in the blue FSRQ case. The location of the emitting region is not constrained to be beyond the BLR, but the jet parameters of this case are very similar to the blue FSRQ. The bottom panel of Figure \ref{combo} shows also the comparison of the SED of 0040--0915 with the SED of PKS 2155--304, if the latter had $z=4.98$. We can see that the current radio upper limit is not severe enough to exclude the BL Lac interpretation, but future deeper radio observations can help to discriminate. 

In conclusion, our modelling, which combines optical and multi-wavelength data with three possible scenarios, cannot rule out the possibility that this source is indeed a BL Lac object or a blue FSRQ. A critical point in this interpretation is that we expect a radio emission with a flux slightly below the current upper limit. Therefore, we cannot establish without any doubt the classification of 0040--0915 yet, and deeper radio observations are mandatory. From the optical point of view a monitoring of the source could reveal variability or flares expected from non-thermal dominated sources.
The newly presented SED  prefers the idea that the optical emission is dominated by the direct emission from an accretion disk as expected from a normal QSO. The low luminosity of the lines is puzzling and makes 0040--0915 an example of high-redshift weak emission line quasar as discussed by many authors in literature (see, e.g. \cite{dia09}).
\\
In order to speculate a little bit more on this final interpretation, we try to accommodate the observed source in the model framework recently drawn in \cite{wu,luo} and further discussed with medium resolution optical observation of WELQ by \cite{plot15} with ESO-XSHOOTER. According to the proposed model (see sketch in Figure 18 of \cite{luo}) the accretion disk of WELQ is geometrically thick and significantly puffed due to high accretion rate (see also \cite{shem}). The puffed portion of the disk blocks X-Rays and the ionizing continuum, produced in the nucleus, from reaching the broad line region, resulting in the observed weak line emission. Various scenarios for WELQ (also for their X-Rays properties) can therefore be explained by varying the angle of view as clearly depicted in \cite{luo}. In the case of 0040--0915 the X-Rays data seems to be consistent (see Figure 5) with those of normal quasars ($\sim$10$^{45}$ erg cm$^{-2}$ s$^{-1}$ $\textrm{\AA}^{-1}$, see e.g.\cite{mac}) . According to Figure 9 in \cite{wu} and Figure 18 in \cite{luo} when the source is viewed through the shielding gas coming from the puffed disk the X-Rays luminosity is significantly  dimmed. Otherwise, a weak-line quasar with X-ray normal luminosity is seen. In the case of 0040--0915 the X-Rays flux does not appear to be absorbed, suggesting that the source is viewed far away from the shielding gas.

\begin{table*}
\centering
\begin{tabular}{lllll lllll lll}
\hline
\hline
Name &$M_{\rm BH}$ &$\Gamma$  &$\theta_{\rm v}$ &$R_{\rm diss}$ &$R_{\rm BLR}$   &$P^\prime_{\rm i}$ &$B$
&$\gamma_{\rm b}$ &$\gamma_{\rm max}$   &$\alpha_X$ &$E_{\rm c}$\\
~[1]       &[2] &[3] &[4] &[5] &[6] &[7] &[8] &[9] &[10] &[11] &[12]    \\       
\hline
Radio--quiet &2e9 &... &... &... &1132 &...     &... &...   &...   &1.2 &130 \\
Blue FSRQ    &1e9   &14  &3   &360 &212  &4e--3 &1.2 &1.5e4 &1e6   &1.2  &150 \\
BL Lac       &...   &14  &3   &240 &...  &3e--3 &1   &1.4e4 &1e6   &1.   &... \\
\hline
\hline
\end{tabular}
\vskip 0.2 true cm
\caption{
Col. [1]: type of source;
Col. [2]: black hole mass in solar mass units;
Col. [3]: bulk Lorentz factor;
Col. [4]: viewing angle (degrees);
Col. [5]: distance of the blob from the black hole in units of $10^{15}$ cm;
Col. [6]: size of the broad line region in units of $10^{15}$ cm; 
Col. [7]: power injected in the blob calculated in the comoving frame, in units of $10^{45}$ erg s$^{-1}$;
Col. [8]: magnetic field in Gauss;
Col. [9] and [10]: minimum and maximum random Lorentz factors of the injected electrons;
Col. [10]: energy spectral index assumed for the X--ray corona;
Col. [11]: cut--off energy of the X--ray spectrum, in keV
[assumed to be $\propto \nu^{-\alpha_X} \exp(-h\nu/E_{\rm c})$].
}
\label{para}
\end{table*}

\begin{table*}
\centering
\begin{tabular}{lllll ll}
\hline
\hline
Name &$\log L_{\rm d}$ &$L_{\rm d}/L_{\rm Edd}$ &$\log P_{\rm r}$ &$\log P_{\rm B}$ 
&$\log P_{\rm e}$ &$\log P_{\rm p}$ \\
~[1] &[2]              &[3]                     &[4]              &[5]              &[6]              &[7]  \\ 
\hline
Radio--quiet &47 &0.45 &...  &...  &... &...  \\
blue FSRQ    &46 &0.03 &45 &45 &44 &44  \\
BL Lac       &...  &...  &45 &45 &44 &44  \\
\hline
\hline
\end{tabular}
\vskip 0.2 true cm
\caption{
Accretion and jet powers:
Col. [1]: type of source;
Col. [2]: logarithm of the accretion disk luminosity (units are erg s$^{-1}$);
Col. [3]: accretion disk luminosity in Eddington units;
Col. [4]--[7]: logarithm of the jet power in the form of radiation ($P_{\rm r}$), Poynting flux ($P_{\rm B}$),  
bulk motion of electrons ($P_{\rm e}$) and [7] protons ($P_{\rm p}$), assuming one cold proton
per emitting electron. Units are erg s$^{-1}$.
These values refer to two jets.
}
\label{powers}
\end{table*}

\acknowledgments

Acknowledgments:
\\
We thank the anonymous referee for his/her careful review of our manuscript that allowed to greatly improve the presentation of our work.We are also grateful to Dr. John Telting, all the technicians and the astronomers at the Nordic Optical Telescope for the support during commissioning and for the acquired observations which this paper refers to. The authors would like to thank also Dr. Thomas F\"{a}cke and Dr. Enrico Orselli (Bayer MaterialScience AG) for providing the custom photopolymeric film for the grating manufacturing. This work was partly supported by the European Community (FP7) through the OPTICON project (Optical Infrared Co-ordination Network for astronomy).

\end{document}